\documentclass[journal=jacsat,manuscript=article,layout=twocolumn]{achemso}
\setkeys{acs}{maxauthors = 10, etalmode=truncate}
\usepackage[version=3]{mhchem} 
\usepackage{tabularx}
\usepackage{amsmath}
\usepackage{amssymb}
\usepackage[utf8x]{inputenc}

\newcolumntype{Y}{>{\centering\arraybackslash}X}
\newcolumntype{s}{>{\hsize=.5\hsize\centering\arraybackslash}X}

\author{Timothy J. Barnum}
\affiliation{Department of Chemistry, Union College, Schenectady, NY 12308}
\email{barnumt@union.edu}

\author{Mark A. Siebert}
\affiliation{Department of Astronomy, University of Virginia, Charlottesville, VA 22904, USA}

\author{Kin Long Kelvin Lee}
\affiliation{Department of Chemistry, Massachusetts Institute of Technology, Cambridge, MA 02139, USA}

\author{Ryan A. Loomis}
\affiliation{National Radio Astronomy Observatory, Charlottesville, VA 22903, USA}

\author{P. Bryan Changala}
\affiliation{Center for Astrophysics $\mid$ Harvard~\&~Smithsonian, Cambridge, MA 02138, USA}

\author{Steven B. Charnley}
\affiliation{Astrochemistry Laboratory and the Goddard Center for Astrobiology, NASA Goddard Space Flight Center, Greenbelt, MD 20771, USA.}

\author{Madelyn L. Sita}
\affiliation{Department of Chemistry, University of Virginia, Charlottesville, VA 22904, USA}

\author{Ci Xue}
\affiliation{Department of Chemistry, Massachusetts Institute of Technology, Cambridge, MA 02139, USA}

\author{ Anthony J. Remijan}
\affiliation{National Radio Astronomy Observatory, Charlottesville, VA 22903, USA}

\author{ Andrew M. Burkhardt}
\affiliation{Department of Physics, Wellesley College, 106 Central Street, Wellesley, MA 02481, U.S.A.}

\author{Brett A. McGuire}
\affiliation{Department of Chemistry, Massachusetts Institute of Technology, Cambridge, MA 02139, USA}
\alsoaffiliation{National Radio Astronomy Observatory, Charlottesville, VA 22903, USA}
\alsoaffiliation{Center for Astrophysics $\mid$ Harvard~\&~Smithsonian, Cambridge, MA 02138, USA}
\email{brettmc@mit.edu}

\author{Ilsa R. Cooke}
\affiliation{Department of Chemistry, University of British Columbia, 2036 Main Mall, Vancouver BC V6T 1Z1}
\email{icooke@chem.ubc.ca}

\title[An \textsf{achemso} demo]
  {A Search for Heterocycles in GOTHAM Observations of TMC-1}

\abbreviations{IR,NMR,UV}

\begin{document}

\begin{tocentry}

\includegraphics[width = 8.25 cm]{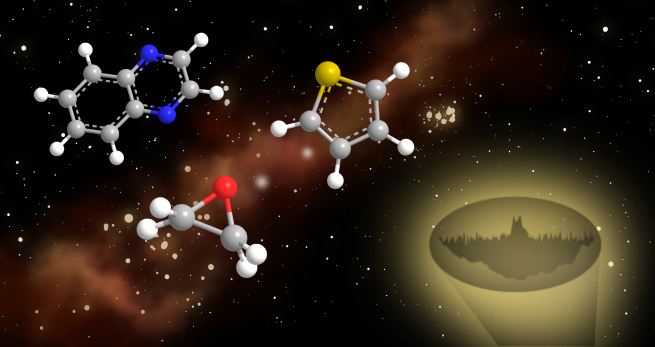}

\end{tocentry}

\begin{abstract}
We have conducted an extensive search for nitrogen-, oxygen- and sulfur-bearing heterocycles toward Taurus Molecular Cloud 1 (TMC-1) using the deep, broadband centimeter-wavelength spectral line survey of the region from the GOTHAM large project on the Green Bank Telescope. Despite their ubiquity in terrestrial chemistry, and the confirmed presence of a number of cyclic and polycyclic hydrocarbon species in the source, we find no evidence for the presence of any heterocyclic species.  Here, we report the derived upper limits on the column densities of these molecules obtained by Markov Chain Monte Carlo (MCMC) analysis and compare this approach to traditional single-line upper limit measurements. We further hypothesize why these molecules are absent in our data, how they might form in interstellar space, and the nature of observations that would be needed to secure their detection. 
\end{abstract}

\section{Introduction}

Aromatic compounds are an important class of molecules of astrophysical and astrobiological interest. In particular, polycyclic aromatic hydrocarbons (PAHs) are abundant in interstellar and circumstellar environments and it has been estimated that as much as 25\% of all interstellar carbon is sequestered in these species.\cite{Tielens:2005ux} Despite their significance to terrestrial chemistry, individual five- and six-membered aromatic rings, as well as PAHs, have only recently been identified in the interstellar medium (ISM).\cite{McGuire:2018it,McCarthy:2021aa,McGuire2021,Burkhardt2021,Lee:2021ud,Cernicharo2021,Cernicharo2021a,Cernicharo2021b}

Heterocycles are a related class of cyclic molecules with atoms of at least two different elements as members of its ring(s). In general, a CH unit in a carbon ring is replaced by a heavier heteroatom, such as N, O, or S. Among these, N-heterocycles are of great biological significance as they form the backbone of nucleobases, that is, sub-units of the DNA and RNA that carry the genetic information of living systems. 

The astrophysical relevance of heterocycles has been appreciated since the discovery of N-, O- and S-heterocycles in carbonaceous chondrites.\cite{Callahan2011} For example, a number of S-heterocycles have been detected in the Yamato-791198 meteorite,\cite{Komiya1993} while N-heterocycles including pyridines, pyrimidines, quinolines, and isoquinolines have been detected in several chondrites, including Orgueil,\cite{Ryoichi1964} Murchison,\cite{FOLSOME1971,Folsome1973,Hayatsu1975,Velden1977} and Murray.\cite{Stoks1979}. Their extraterrestrial origins have been confirmed by isotopic studies,\cite{Martins2008} but it remains difficult to ascertain their chemical origin.\cite{Martins2018} 

Following the recent detection of benzonitrile (c-C$_6$H$_5$CN) in the quiescent molecular cloud Taurus Molecular Cloud 1 (TMC-1),\cite{McGuire:2018it} both five-membered and bicyclic nitrile species (hydrocarbons with a cyanide [\ce{-CN}] functional group) have been detected using our GOTHAM line survey\cite{McCarthy:2021aa,McGuire2021,Burkhardt2021,Lee:2021ud} and the QUIJOTE survey of Cernicharo and colleagues.\cite{Cernicharo2021,Cernicharo2021a,Cernicharo2021b} Benzonitrile has also been observed in four additional molecular clouds in different stages of prestellar evolution, suggesting that aromatic molecules are ubiquitous in interstellar environments.\cite{Burkhardt:2021aa} 

While these detections have shown that aromatic molecules are prevalent in TMC-1, these molecules are either pure aromatic hydrocarbons or functionalized aromatics with a \ce{-CN} or C$_2$H group in the place of a hydrogen on the ring.  With the detection of several aromatics now secure, heterocyclic molecules are the next logical step in chemical complexity to explore. Yet, to date no five- or six-membered ring heterocycles have been detected in TMC-1 nor in any region of interstellar space.\cite{McGuire:2022wl} For these relatively large molecules, the number of possible isomers renders prioritization difficult, and thus we choose to combine conventional chemical intuition and recent machine learning work by \citet{leeMachineLearningInterstellar2021}.

\begin{figure*}
    \centering
    \includegraphics[width=\textwidth]{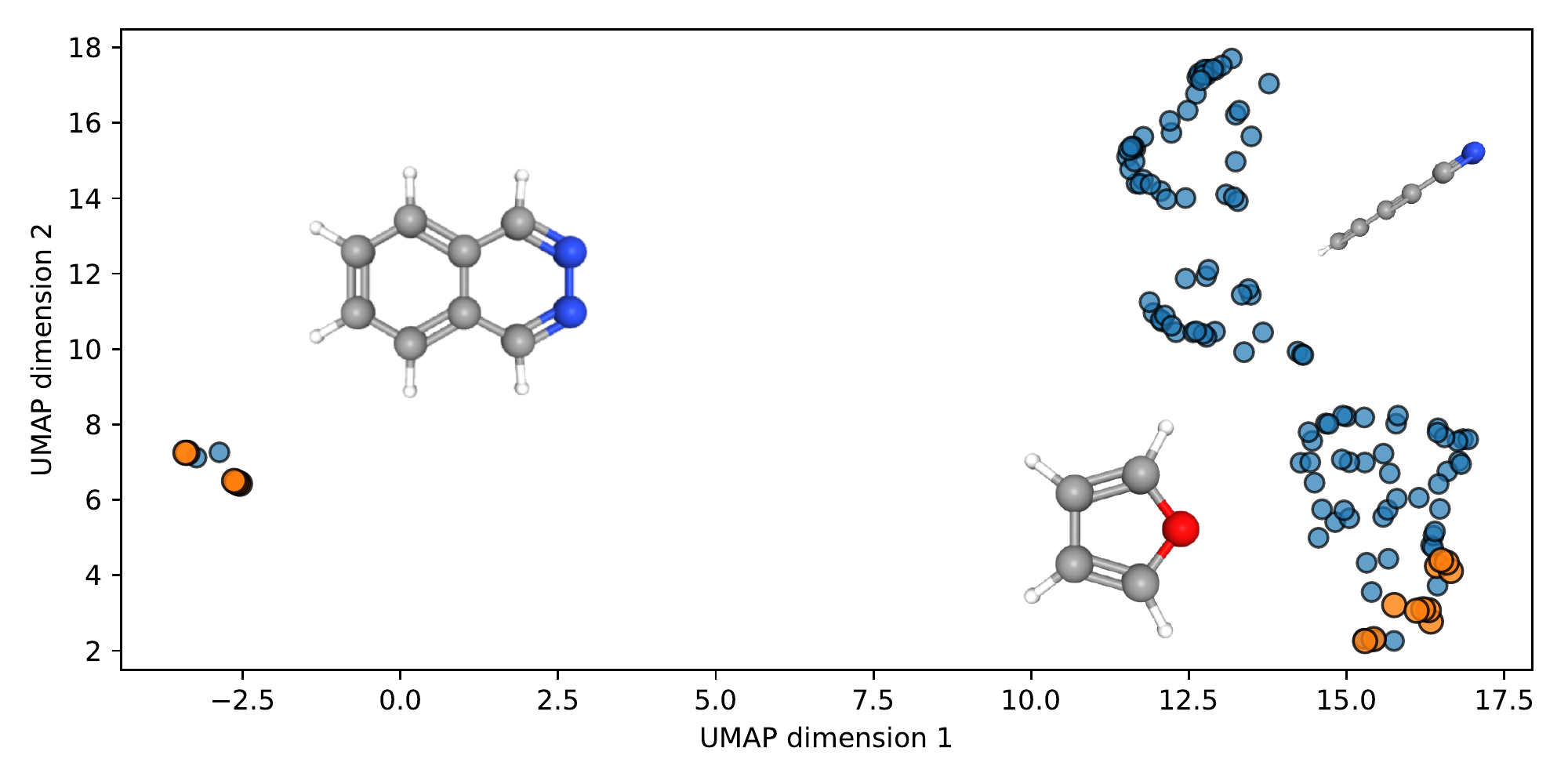}
    \caption{Visualization of a two-dimensional projection of the chemical embedding space spanned by: the current chemical inventory of TMC-1 (blue); and the proposed heterocyclic species (orange). Representative structures of indicative molecules annotate how chemical complexity in this source is traversed and divided: molecules such as the cyanopolyynes extend from the top to the bottom on the right, connecting with single-ring heterocycles such as furan toward the bottom. On the left, two-ring species such as cyanonaphthalene (not pictured) and their heterocyclic analogues.}
    \label{fig:umap}
\end{figure*}

To illustrate how machine learning can be used to guide astrochemical studies, Figure \ref{fig:umap} represents a ``chemical map'', where the abscissa and ordinate represent the 2D space that comprises molecules found in TMC-1 and those studied in the present work; the absolute values are not important, however the relative distance between points (i.e. molecules) corresponds with chemical similarity. To briefly summarize the technique, the Uniform Manifold Approximation and Projection (UMAP) method \cite{umap-viz} is an unsupervised algorithm that learns an approximate mapping between high dimensional data---in this case, a 128-dimensional chemical embedding obtained through natural language models \cite{lee_kin_long_kelvin_2021_6299773,leeMachineLearningInterstellar2021}---to lower dimensionality whilst preserving the topology of the original manifold as faithfully as possible. The target species comprise two frontiers: monocyclic (lower right of Fig. \ref{fig:umap}) and bicyclic (left of Fig. \ref{fig:umap}) rings, which are currently sparsely represented and for that reason, are among the most informative to quantify in TMC-1. By constraining the abundance of these species, we are able to effectively and precisely place boundary conditions on one edge of chemical complexity for an exemplary, molecule-rich dark molecular cloud.

\subsection{Prior searches}

Previous searches have predominantly focused on nitrogen-bearing heterocycles, with the only exceptions being the O- and Si-bearing species oxirane,\cite{Dickens1997} propylene oxide,\cite{Mcguire2016}  furan,\cite{Kutner1980,Dickens2001,Barnum2021} and silicon di- and tri-carbides.\cite{Thaddeus1984,Apponi1999} To the best of our knowledge, there have not been any searches for sulfur-bearing heterocyles. The emphasis on N-heterocycles is likely motivated by their significance in biological molecules, combined with the availability of spectroscopic data. Below we outline the prior searches for nitrogen and oxygen bearing heterocycles that have been conducted. For completeness, we note that two silicon bearing heterocycles have been detected toward IRC+10216, silicon dicarbide (silacyclopropynylidene, SiC$_2$)\cite{Thaddeus1984} and silicon tricarbide (SiC$_3$).\cite{Apponi1999}

\subsubsection{Detections}

\textbf{Oxirane, c-C$_2$H$_4$O:} Oxirane (ethylene oxide) is the simplest oxygen-bearing heterocycle. It was first detected by Dickens et al. using lines from the Nobeyama 45-m, Haystack 140-ft, and SEST 15-m telescope observed toward Sagittarius B2 (Sgr B2).\cite{Dickens1997} It has since been detected in various interstellar environments including low mass protostars\cite{Lykke2017} and prestellar cores.\cite{Bacmann2019}\\
 
\noindent\textbf{2-Methyloxirane, c-CH$_3$C$_2$H$_2$O:} The functionalized heterocycle, 2-methyloxirane (propylene oxide) was detected by McGuire et al. in GBT 100-m and Parkes 64-m observations of Sgr B2.\cite{Mcguire2016} 

\subsubsection{Tentative detections}

\textbf{$2H$-azirine, C$_2$H$_3$N:} $2H$-azirine, c-C$_2$H$_3$N, and its reduced form aziridine (ethylenimine), c-C$_2$H$_5$N, are the simplest N-heterocyclic compounds. $2H$-azirine has been previously searched for in Sgr B2(N-LMH) and Orion KL and a tentative detection was reported.\cite{Kuan2003,Kuan2004,Charnley2001}

\noindent\textbf{Aziridine, C$_2$H$_5$N:} Two groups have reported tentative detections of aziridine. \citet{Dickens2001} reported two detected lines toward hot cores G327.3, G10.47, and possible weak emissions toward G34.3, Sgr B2(N) and NGC6334. Kuan et al. also reported a tentative detection of aziridine toward hot cores Orion KL and W51 e1/e2.\cite{Kuan2003,Kuan2004}
\subsubsection{Non-detections}

\textbf{Pyrrole, C$_4$H$_4$NH:} Pyrrole, or more specifically $1H$-pyrrole, is a five-membered ring analogous to cyclopentadiene, in which a carbon is substituted for a nitrogen atom. Two other tautomers of pyrrole exit:  $2H$-pyrrole, which has the double bonds at positions 1 and 3 and 3H-pyrrole, which has the double bonds at positions 1 and 4. Pyrrole has been searched for toward the hot core Sgr B2(N)\cite{Myers1980,Kutner1980} as well as the cold molecular cloud TMC-1\cite{Kutner1980}.  Toward TMC-1, Kutner et al. derived an upper limit of 4$\times$10$^{12}$ cm$^{-2}$.

\noindent\textbf{Imidazole, C$_3$N$_2$H$_4$:} Imidazole is a diazole (five-membered aromatic heterocycle with two nitrogen atoms) of particular biological relevance. This ring system is present in important biological building blocks, such as histidine and the related hormone histamine. In addition, many pharmaceutical drugs contain an imidazole ring.\cite{Zhang2013} When fused to a pyrimidine ring, it forms a purine, which is the most widely occurring nitrogen-containing heterocycle in nature.\cite{Rosemeyer2004} Despite its relevance on earth, imidazole has only been searched for toward Sgr A and Sgr B2(N).\cite{Dezafra1972,Giuliano2019}  \citet{Giuliano2019} recently searched for imidazole in Sgr B2(N), following their improved laboratory spectroscopy, and reported a non-detection. 
\noindent\textbf{Pyridine, C$_5$H$_5$N:} Pyridine is a 6-membered heterocycle, structurally related to benzene, with one CH replaced by a nitrogen atom. The pyridine ring occurs in many important compounds, including agrochemicals, pharmaceuticals, and vitamins.\cite{Vitaku2014}
Pyridine has been searched for toward hot cores Sgr B2 \cite{Simon1973,Batchelor1973}, Sgr A,\cite{Batchelor1973} Orion A,\cite{Batchelor1973}, and the circumstellar envelopes of carbon-rich stars, IRC+10216 and CRL 618.\cite{Charnley2005}

\noindent\textbf{Pyrimidine, C$_4$H$_4$N$_2$:} Pyrimidine is an aromatic heterocycle similar to pyridine. One of the three diazines (six-membered aromatic heterocycles with two nitrogen atoms), it has the nitrogen atoms at positions 1 and 3 in the ring. The other diazines are pyrazine (nitrogen atoms at the 1 and 4 positions) and pyridazine (nitrogen atoms at the 1 and 2 positions), which have not been previously searched for in the ISM. Pyrimidine occurs widely in nature, including in the nucleotides cytosine, thymine and uracil, thiamine (vitamin B1) and alloxan, as well as in many synthetic drugs.\cite{Jeelan2021} Pyrimidine was first searched for in 1973 by Simon and Simon toward the hot core Sgr B2(N).\cite{Simon1973} Following the non-detection in Sgr B2(N), \citet{Irvine1981} searched for pyrimidine in a handful of sources: TMC-1, L134N, W3 C, W3 (OH), Orion A, DR 21 (OH), S140 and IRC+10216. They reported an upper limit  of 6$\times$10$^{12}$ cm$^{-2}$ toward TMC-1. Kuan et al. later conducted a search for pyrimidine toward Sgr B2(N), Orion KL and W51 e1/e1. 

\noindent\textbf{Quinoline, C$_9$H$_7$N:} Quinoline is one of the simplest polycyclic aromatic nitrogen heterocycles (PANHs), with a similar structure to naphthalene but with one CH substituted for a nitrogen. Quinoline was searched for by \citet{Charnley2005} toward the circumstellar envelopes IRC+10216, CRL 618 and CRL 2688. 

\noindent\textbf{Isoquinoline, $i$-C$_9$H$_7$N:} Isoquinoline is a structural isomer of quinoline, with CH  substituted for a nitrogen atom at the 2 position in the ring. Isoquinoline was likewise searched for by Charnley et al. toward the circumstellar envelopes IRC+10216, CRL 618 and CRL 2688.\cite{Charnley2005}
assumed. 

\noindent\textbf{Furan, C$_4$H$_4$O:} Furan is a five-membered aromatic heterocycle with four carbon atoms and one oxygen. It was first searched for by \citet{Kutner1980} toward hot cores Sgr B2(N) and Orion A. \citet{Dickens2001} searched for furan toward a number of sources: TMC-1(CP), Orion KL, Orion 3'N, G327.3-0.6, L134N(C), IRS 16293, NGC 6334F, Sgr B2(N), and W33A. Toward TMC-1(CP) they reported an upper limit of 1--2$\times$10$^{13}$ cm$^{-2}$. Most recently, \citet{Barnum2021} reported an upper limit of 1$\times$10$^{12}$ cm$^{-2}$ using the GOTHAM line survey.

\begin{figure*}
    \centering
    \includegraphics[width=0.8\textwidth]{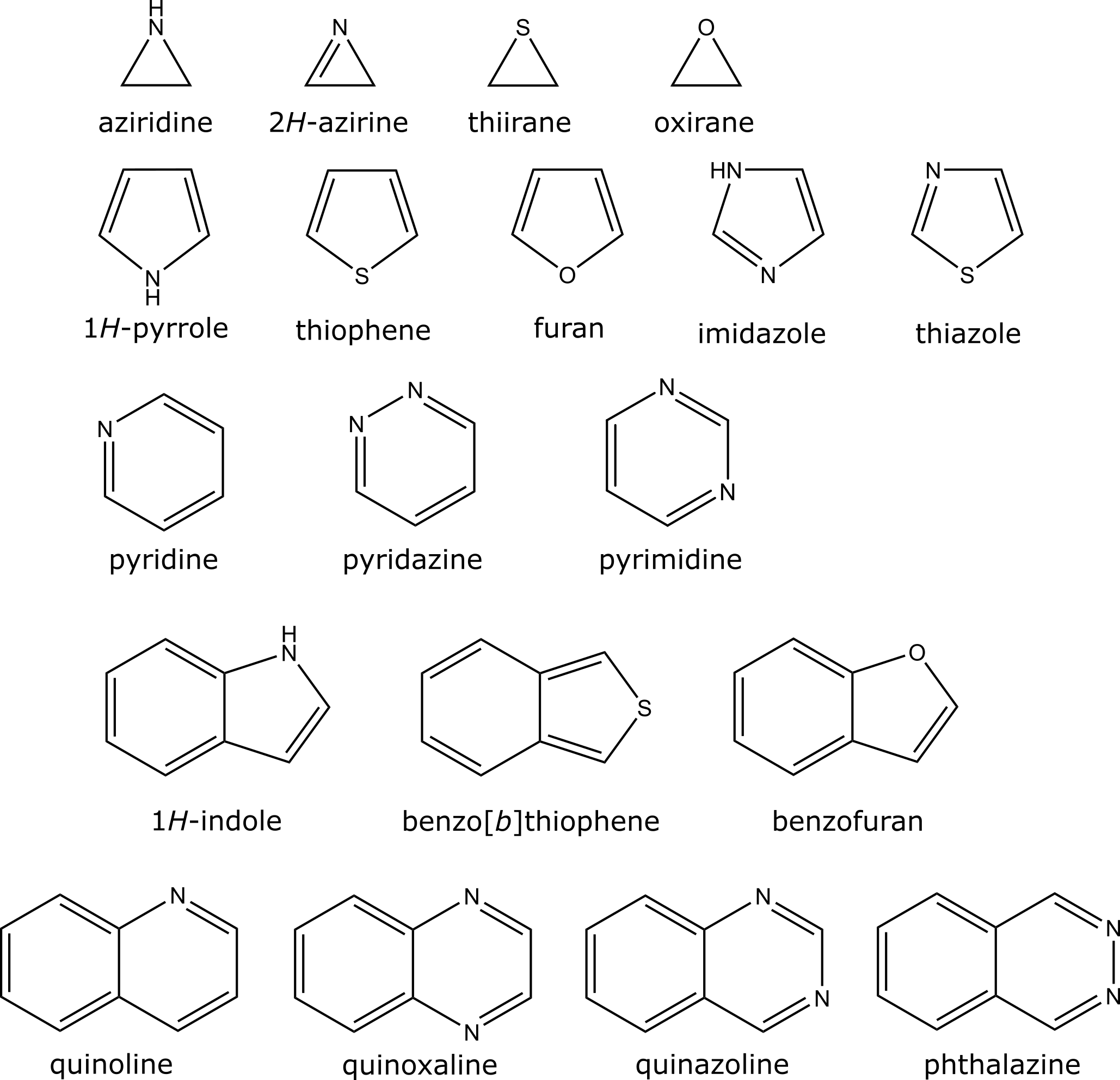}
    \caption{Chemical structures of the heterocycles included in our analysis.}
    \label{fig:structures}
\end{figure*}

\section{Methods}
\subsection{Observations}
\label{sec:obs}
We performed a search toward TMC-1 using the third data release of the GOTHAM collaboration survey. Details of the source, observations, and data reduction methods can be found in \citet{McGuire:2020bb} and \citet{McGuire2021} Here, we provide a brief summary. 
Observations were performed with the 100-m Robert C. Byrd Green Bank Telescope 
with the project codes GBT17A\nobreakdash-164,  GBT17A\nobreakdash-434, GBT18A\nobreakdash-333, GBT18B\nobreakdash-007, and data from project GBT19B\nobreakdash-047 acquired through April 2021.

The first and second data release of GOTHAM (hereafter referred as DR1 and DR2) comprise observations obtained between February 2018 -- May 2019 (DR1) and until June 2020 (DR2).\cite{McGuire:2020bb,McGuire2021} The GOTHAM observations used here are the third data reduction (hereafter referred to as DR3), which comprises observations made through April 2021. DR3 extends the frequency coverage to 7.906 -- 35.434 GHz with a few gaps and improved the sensitivity in some frequency coverage already covered by DR2.

The pointing was centered on TMC\nobreakdash-1 CP at (J2000) $\alpha$~=~04$^h$41$^m$42.50$^s$ $\delta$~=~+25$^{\circ}$41$^{\prime}$26.8$^{\prime\prime}$. The spectra were obtained through position-switching to an emission-free position 1$^{\circ}$ away. Pointing and focusing was refined every 1--2 hours on the calibrator J0530+1331. Flux calibration was performed with an internal noise diode and Very Large Array (VLA) observations (Project TCAL0003) of the same calibrator used for pointing, resulting in a flux uncertainty of ${\sim}20$\%. \citep{McGuire:2020bb}

Our search includes 19 heterocycles whose chemical structures are shown in Figure \ref{fig:structures}. The molecules chosen include small, single-ring heterocycles, as well as two-ring species and those with two heteroatoms in the ring. Factors influencing our choice of molecules included chemical complexity (i.e. starting with small, single-heteroatom species), prior searches, the availability of spectroscopic data, and the recommendations produced by the unsupervised machine learning model of \citet{Lee2021} Our search is heavily biased toward nitrogen heterocycles, which is perhaps not surprising given the prevalence of nitrogen-containing molecules already detected in TMC-1, their dominance in previous searches, and their biological relevance. 

Figure \ref{fig:umap} shows a visualization of the chemical space we have chosen to prioritize for this work: many heterocycles were not explored, including some unsubstituted heterocycles for which sufficient spectroscopic data could not be sourced. This list includes the four-membered rings azete, $2H$-oxete and $2H$-thiete (and their saturated counterparts azetidine, oxetane and thietane); 6-membered rings containing sulfur or oxygen (pyrans and thiopyrans), as well as oxygen and sulfur containing polycyclic heterocyles (chromene, thiochromene). Polycylic heteocycles containing more than one heteroatom were likewise not considered in this study. Structural isomers and/or tautomers may also be considered in future studies, for example $2H$-pyrrole, isobenzofuran, isoindole and isoquinoline.

\section{Results and Discussion}
\begin{table*}
    \caption{Upper limit column densities for all investigated heterocycles determined as the 97.5th percentile of the highest posterior density credible interval summed over all four components of TMC-1. The upper limits are reported as log column densities in units of cm$^{-2}$. The table also reports the largest dipole moment component in the principal axis system of each molecule in units of Debye, and the total number of lines included in the analysis.}
    \label{table:upperlimits}
    \begin{tabularx}{\linewidth}{Y s Y *3s}
    \hline
species & formula & \parbox[t]{\hsize}{\centering column density upper limit \\ (log$_{10}$ cm$^{-2}$)}  & dipole (D) & lines & refs \\ 
 \hline
 aziridine & \ce{C2H5N} & 13.29 & 1.36 & 155 & ~\citenum{Thorwirth2000} \\
 2H-azirine & \ce{C2H3N} & 12.10 & 2.07 & 34 & ~\citenum{Bogey1986} \\
 thiirane & \ce{C2H4S} & 12.33 & 1.84 & 46 & ~\citenum{Hirao2001}\\
 oxirane & \ce{C2H4O} & 12.26 & 1.88 & 6 & ~\citenum{Medcraft2012,Mueller2022} \\
 1H-pyrrole & \ce{C4H5N} & 12.00 & 1.74 & 382 & ~\citenum{Wlodarczak1988,Nygaard1969}\\
 thiophene & \ce{C4H4S} & 13.10 & 0.55 & 68 & ~\citenum{Orr2021} \\
 furan & \ce{C4H4O} & 12.93 & 0.66 & 103 & ~\citenum{Barnum2021} \\
 imidazole & \ce{C3H4N2} & 12.09 & 3.60 & 1456 & ~\citenum{Giuliano2019}\\
 thiazole & \ce{C3H3NS} & 12.85 & 1.29 & 372 & ~\citenum{Esselman2021,Kretschmer1993}\\
 pyridine & \ce{C5H5N} & 11.88 & 2.22 & 775 & ~\citenum{McCarthy2020}\\
 pyridazine & \ce{C4H4N2} & 12.33 & 4.22 & 1503 & ~\citenum{Esselman2013,Werner1967}\\
 pyrimidine & \ce{C4H4N2} & 12.62 & 2.33 & 659 & ~\citenum{Heim2020,Kisiel1999,Blackman1970} \\
 1H-indole & \ce{C8H7N} & 11.60 & 1.59 & 4203 & ~\citenum{Nesvadba2017}\\
 benzo[b]thiophene & \ce{C8H6S} & 13.11 & 0.64 & 1562 & ~\citenum{Welzel1999} \\
 1-benzofuran & \ce{C8H6O} & 12.01 & 0.72 & 1147 & ~\citenum{Maris2005}\\
 quinoline & \ce{C9H7N} & 12.38 & 2.01 & 9022 & ~\citenum{Kisiel2003}\\
 quinoxaline & \ce{C8H6N2} & 13.06 & 0.59 & 1128 & ~\citenum{McNaughton2011}\\
 quinazoline & \ce{C8H6N2} & 10.61 & 2.89 & 10487 & ~\citenum{McNaughton2011} \\
 phthalazine & \ce{C8H6N2} & 10.94 & 5.45 & 5189 &~\citenum{McNaughton2011}\\ 
    \hline
    \end{tabularx}
\end{table*}

\begin{figure}
    \centering
    \includegraphics[width=\columnwidth]{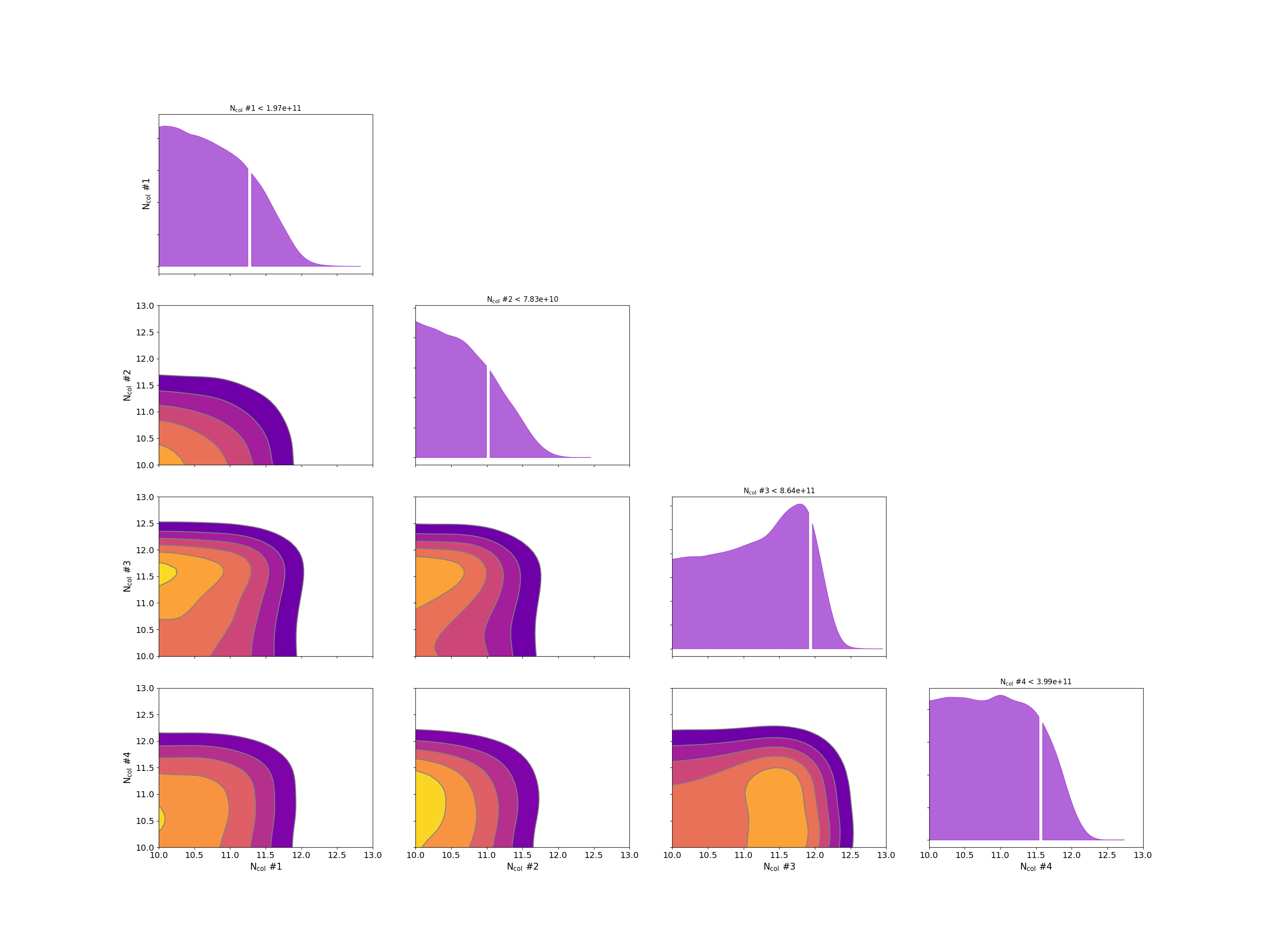}
    \caption{Corner plot of the marginal posterior distribution of the four column densities for 1H-pyrrole. Tick marks on all axes represent orders of magnitude of the column density in units of cm$^2$. In all four velocity components, a long-tailed distribution toward low column densities indicate a non-detection of this species in the data. The derived upper limit is indicated by the gap in the marginalized 1-D distribution for each component and the upper limit value appears above.}
    \label{fig:pyrrole_cornerplot}
\end{figure}

\subsection{Upper Limits Analysis}

In the third data release from the GOTHAM collaboration, we could not identify individual lines belonging to the heterocycles in Table \ref{table:upperlimits}. Without individual line detections, we applied Bayesian forward modeling with Markov chain Monte Carlo (MCMC) sampling to establish upper limits to the column density of these molecules in TMC-1. These methods have been described in detail previously \citep{Loomis:2020io} and have been used in combination with spectral line stacking to positively identify new species in TMC-1 without individual observable transitions.

These methods are implemented in \texttt{molsim}, \citep{lee_kin_long_kelvin_2021_5497790} wrapping the affine-invariant sampler \texttt{emcee}\cite{ForemanMackey2013} and uses \texttt{ArviZ}\cite{Kumar2019} for analysis and visualization. Here, we encode knowledge about the physical conditions of TMC-1 and the heterocyclic species through the choice of prior distributions. For all species, we used Gaussian priors consistent with the posterior distributions derived from our earlier analysis of benzonitrile,\citep{McGuire:2020bb} which includes four velocity components with individual source sizes and radial velocities, as well as a shared excitation temperature and linewidth parameter. The excitation temperature prior was adjusted upwards from that earlier analysis to a mean value of $T_{ex}=8.0$ K, in better agreement with more recent analyses in TMC-1 by both the GOTHAM and QUIJOTE projects.\citep{Burkhardt:2021, Shingledecker:2021, Agundez:2022} Uniform priors spanning 16 orders of magnitude for the four column densities were used to minimize the influence of the prior choice on the derived column density upper limits. In all cases, sampling of 200 Markov chains with 10000 steps led to acceptable convergence ($\hat{r} \lesssim 1.1$) of all parameters. In Table \ref{table:upperlimits}, we report 2$\sigma$ upper limit log column densities for all species, corresponding to the 97.5th percentile of the highest density credible interval of the resultant posterior distributions. A representative corner plot of the column density posterior distributions for the species $1H$-pyrrole appears in Figure \ref{fig:pyrrole_cornerplot}. The diagonal traces correspond with the marginalized likelihood for each column density, while the off-diagonal contour plots show the covariance between each parameter. The upper limit is indicated by the break in the marginalized 1-D distributions, and its value appears above each plot. The long-tailed distributions toward zero column densities are characteristic of posterior distributions representing upper limits.

For those species previously searched for in TMC-1 ($1H$-pyrrole, pyrimidine, furan), \citep{Kutner1980,Irvine1981,Dickens2001} we find that our data provides a more stringent upper limit. In the case of furan, which was previously searched for in TMC-1 by some of us, \cite{Barnum2021} the upper limit presented here represents a more conservative determination. The difference between our previous and current values is attributed to a difference in assumptions between the two analyses. Here, we allow the source size to vary and use uniform column density priors, rather than fixing the source sizes and using Gaussian column density priors.

To facilitate comparison with other upper limit determinations in the literature, we also derived upper limits for four heterocycles using a more traditional, frequentist single-line analysis. For each molecule, we used \texttt{molsim} to simulate a spectrum at the column density and excitation conditions of benzonitrile ($T_{ex}=8$\,K) using a single velocity component with a width of 0.5\,km/s at $v_{lsr}=5.85$\,km/s.\citep{Kaifu:2004tk} We then identified the brightest line covered by our spectra, prioritizing transitions that lie in regions where the root-mean-square (RMS) noise is low, and applied the formalism of \citet{Hollis:2006uc} to determine the upper limit column density.  

Here, $T_{ex}$ is the assumed 8\,K excitation temperature (K); $T_{bg}$ is the background continuum temperature and assumed to be 2.7~K; $k$ is Boltzmann's constant (J~K$^{-1}$); $h$ is Planck's constant (J~s); $Q$ is the rotational partition function at $T_{ex}$; $E_{u}$ is the upper state energy of the transition (K); $\int TdV$ is the upper limit velocity-integrated line area (K $\cdot$ km/s), which is taken as the product of the $2\sigma$ rms noise level and the assumed 0.5\,km/s line width; $S_{ij}\mu^2$ is the transition line strength (Debye$^2$); $B$ and $\eta_B$ are the beam filling factor and beam efficiency, which are both assumed to be 1.0 for this analysis.

\begin{equation}
    N_T=  \frac{1}{2}  \frac{3k}{8\pi^3}  \sqrt{\frac{\pi}{\ln 2}}\frac{Qe^\frac{E_u}{T_{ex}}\int TdV}{B\nu S_{ij}\mu^2\eta_B} \frac{1}{1-\frac{e^{\frac{h\nu}{kT_{ex}}{-1}}}{e^{\frac{h\nu}{kT_{bg}}{-1}}}}
\end{equation}

\begin{figure}
    \centering
    \includegraphics[width=\columnwidth]{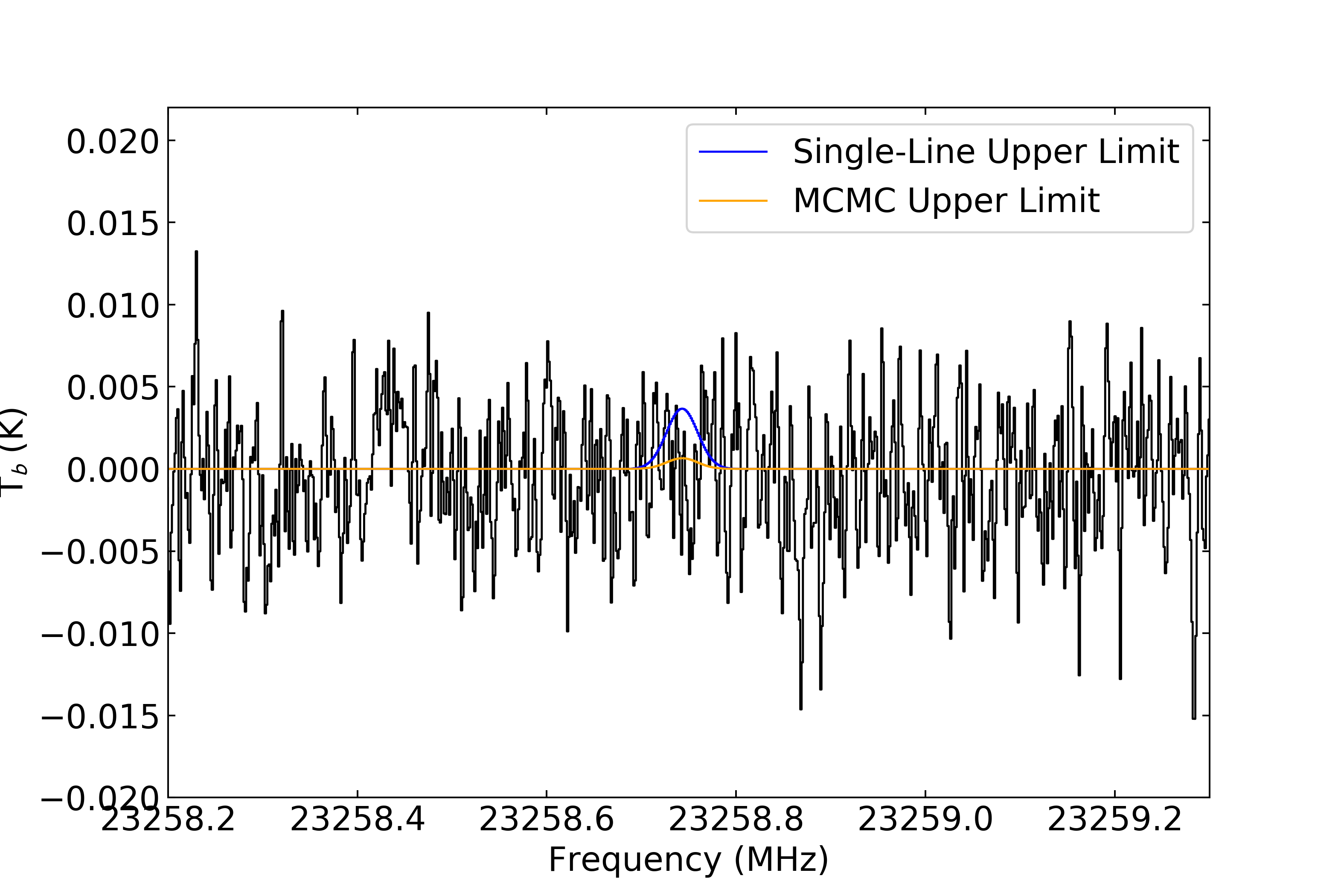}
    \caption{GOTHAM DR3 spectrum (black trace) at the brightest line expected for furan ($J_{K_a, K_c}=2_{1,2} \rightarrow 1_{1,1}$). Simulations plotted in blue and orange correspond to the upper limit column densities derived using the single-line and Bayesian stacking analyses employed in this work. Both simulations were made with $T_{ex}=8$\,K and $v_{lsr}=5.85$\,km/s.}
    \label{fig:furan_comp}
\end{figure}

For furan, the brightest expected line is the $J_{K_a, K_c}=2_{1,2} \rightarrow 1_{1,1}$ transition at 23259.2\,MHz (See Fig. \ref{fig:furan_comp}). Using the sensitivity of our survey at this frequency, we calculate an upper limit log column density of 13.68\,cm$^{-2}$ for furan based on this line. The order of magnitude difference between this approach and the MCMC analysis can be attributed to our consideration of the full spectrum. For furan, 103 individual lines were used in the MCMC analysis, after removing any lines from the analysis that appear near interloping lines of another species within a $\pm 0.5$ m/s window in velocity space. Under the assumptions that all transitions are all similarly bright, and that we only have white noise present in our data, we would expect the RMS of a stacked spectrum containing all of these lines (and thereby, the derived upper limit column density) to go approximately as $1/\sqrt{n}$ where $n$ is the number of stacked lines. The single line analysis of furan is representative of this behavior.

When we apply this method to imidazole, we find that a bright blend of three hyperfine components of the  $J_{K_a, K_c}=3_{1,3} \rightarrow 2_{1,2}$ transition is expected at 33397.38\,MHz. Using this single feature, we derive an upper limit log column density of 12.86\,cm$^{-2}$ for imidazole, which is almost one order of magnitude larger than the upper limit we place using the MCMC approach. In contrast to furan however, a total of 1456 lines were used in the MCMC analysis, so this does not follow the $1/\sqrt{n}$ scaling law discussed above. This is likely due to the inclusion of hyperfine structure in our catalog for this molecule, which often produces blended lines that contribute the same information despite being counted as individual lines. For quinazoline, we calculate a log upper limit of 12.50\,cm$^{-2}$ based on 9 blended lines of the $J_{K_a, K_c}=14_{0,14} \rightarrow 13_{0,13}$ at 26536.73\,MHz. This is almost two orders of magnitude larger than what we obtain from a Bayesian analysis of all 10487 covered transitions, meaning a large fraction of those transitions contribute unique information to the MCMC analysis. Finally, for $1H$-pyrrole, we analyzed the $J_{K_a, K_c}=3_{0,3,4} \rightarrow 2_{0,2,3}$ line at 31727.38 MHz to obtain a log upper limit column density of 13.43\,cm$^{-2}$.

As expected, we found that all upper limits derived using a frequentist single-line analysis were larger than those from the MCMC results. The degree by which these upper limits varies is largely determined by the number of lines included in the analysis, the fraction of lines that are not blended, and the fraction that are expected to have significant emission at an excitation temperature of 8\,K. 

\begin{figure}
    \centering
    \includegraphics[scale=0.7]{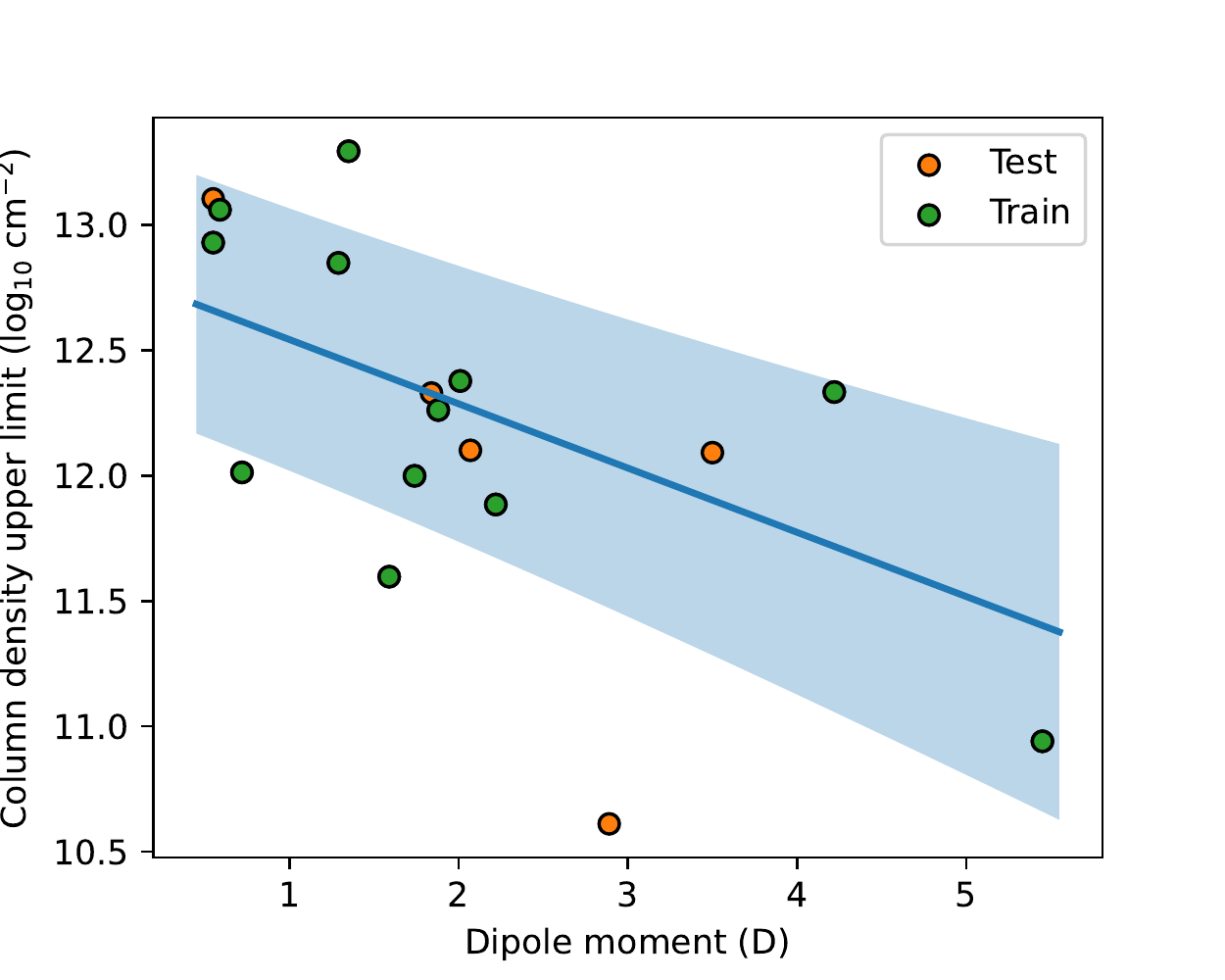}
    \caption{Relationship between molecular properties (dipole moment) and derived upper limit. The blue line and shaded band corresponds to the mean and $\pm1\sigma$ uncertainty of a Bayesian ridge regression model. Scatter points are color coded based on their training/test split.}
    \label{fig:dipoles}
\end{figure}

To describe the relationship between physical properties of molecules and their derived upper limit, Figure \ref{fig:dipoles} plots the column density upper limits as a function of the largest dipole moment component in the principal axis system of each molecule. The fitted model shows that with few exceptions, the upper limit determinations for heterocycles are a function of their physical properties: barring abnormal chemical dynamics (e.g. preferred reactions), one can predict the expected upper limit of other heterocycles bounded by a credible interval, simply from the dipole moment.

We can compare the upper limits of the heterocycles to the derived column densities for the carbocyclic molecules that have been detected in TMC-1 thus far. Pyridine is found to be depleted compared to benzonitrile,\cite{Burkhardt2021} its closest detected carbocyclic counterpart (log column density of 11.88 versus 12.24), suggesting that pyridine is at most half as abundant as benzonitrile. We would therefore expect the CN-substituted pyridine isomers to be substantially lower in abundance than benzonitrile. In the case of the 5-membered rings, cyclopentadiene has been detected recently by \citet{Cernicharo2021} who derived a log column density of 13.08. This suggests that $1H$-pyrrole, thiophene and furan are at least 12, 9.5 and 1.4 times less abundant than their carbocyclic counterpart, further confirming the depletion of heterocycles in TMC-1. Another interesting comparison is between the N-, S- and O- substituted counterparts of indene: indole, benzo$[b]$thiophene and 1-benzofuran. Indole and benzofuran are at least $\sim$20 times and 10 times less abundant than indene, respectively; while the upper limit for benzo$[b]$thiophene is similar to the detected abundance of indene.\cite{Burkhardt2021} As discussed above, molecules with low dipole moments generally have less well constrained upper limits, partially explaining the higher upper limit of benzo$[b]$thiophene. Our results suggest a significant depletion of heterocycles relative to pure carbocycles in TMC-1.

It is noteworthy that for several of the species considered, we observe a distinct peakedness to the column density posterior distributions for one or more velocity components. Manual inspection of the spectral windows used in the MCMC analysis confirms the absence of interloper lines from other detectable species, which could be responsible for these peaked distributions. Instead, we hypothesize that correlated (i.e. ``red") noise in the frequency spectrum is the origin of these unexpected posterior distributions. In the calculation of the log likelihood function, our current MCMC implementation assumes only white noise in the data set, namely that the measurement uncertainty of any single channel in the frequency spectrum is uncorrelated with noise in a neighboring channel. While this assumption captures most of the noise in the astronomical data, it is a first approximation to the noise only. As a result of aperiodic fluctuations in the source brightness over the course of the observational time, red ($\frac{1}{f^2}$) noise also contributes to the noise floor. Beyond intrinsic sources of noise, polynomial baseline fitting in the data reduction step has the potential to introduce additional correlations between channels in the frequency spectrum. Neglecting these correlations in the noise can introduce bias into our inferences, especially as the signals of interest approach and go below the noise floor. For this reason, we do not interpret any peaks in the column density posterior distributions as meaningful and instead interpret all column densities as upper limits only. In a future implementation, we plan to increase the sophistication of our MCMC analysis with noise modeling. By explicit consideration of the off-diagonal components of the covariance matrix used to calculate the likelihood function, we can account for correlated noise in the data set and begin to peer below the noise floor of the GOTHAM data.

\subsection{Discussion on the formation and stability of heterocycles in cold molecular clouds}

The quantitative visualization of chemical space in Figure \ref{fig:umap} underscores the structural similarities between the unobserved heterocycles and previously observed molecules in TMC-1. While the heterocycles do overlap with newly observed aromatic carbocycles, they cluster toward the boundaries of known chemical space in TMC-1. Despite their structural similarity, our null detections for a wide range of heterocycles points to a unique chemistry of these species relative to the pure carbocylic species, which suggests chemically defined bounds on the complexity of molecules found in TMC-1. Given that these heterocycles are highly stable terrestrially, it is likely a matter of dynamics that dictates their absence, and lends importance to finding kinetically viable pathways to their formation.

The formation of heterocycles in TMC-1 would require either chemical pathways that can operate at temperatures below 10 K or their inheritance from high temperature sources, such as circumstellar envelopes. While a number of aromatic molecules have now been detected in TMC-1, it remains unclear whether these molecules form in situ by bottom-up mechanisms from small precursors, or by top-down mechanisms through the destruction of large PAHs or interstellar dust grains.\cite{Burkhardt2021} It is expected that PAHs comprised of less than $\sim$20--30 atoms do not survive transport from the diffuse ISM to cold molecular clouds due to their degradation by ultraviolet photons.\citep{Chabot:2019fw}  While work from the combustion chemistry community has shown that high temperature routes can produce PAHs and nitrogen and oxygen substituted PAHs (PANHs and PAOHs), several low temperature routes have been proposed that may operate in TMC-1. Below, we summarize potential formation routes to heterocycles that could operate under the conditions of TMC-1.

\subsubsection{Formation of the first heterocyclic rings}

Single-ring heterocycles are expected to be building blocks for larger N-, O- and S-substituted PAHs, thus understanding their formation is the first step in determining whether biological precursor heterocycles can exist at the early stages of star-formation.  Formation routes to the smallest heterocycles, aziridine, $2H$-azirine, oxirane, and thiirane, have not been well explored. Crossed-beam experiments suggest the reaction of N($^2$D) with ethylene (C$_2$H$_4$) can form $2H$-azirine; however, isomerization to the more stable isomer acetonitrile (CH$_3$CN) is expected, even under collision-free conditions.\cite{Balucani2000} Recent work has shown the formation of $2H$-azirine in interstellar ice analogues of acetylene and ammonia after irradiation with energetic electrons.\cite{Turner2021} The suggested mechanism involves the reaction of imidogen (NH(a$^1\Delta$)) with acetylene through addition to the carbon-carbon triple bond leading to $1H$-azirine, which may isomerize via hydrogen shift to $2H$-azirine via a barrier of 108 kJ mol$^{-1}$.

Oxirane formation has been explored in more detail, motivated by its astronomical detection in the ISM.\cite{Occhiogrosso2014} Similar to azirine, oxirane has been shown to form in electron\cite{Bennett2005} and photon\cite{Bergner2019} irradiated ices, this time containing carbon dioxide (a source of suprathermal O atoms) and ethylene. Oxirane has also been shown to form ethylene oxide during the co-accretion of thermalized atomic oxygen and ethylene onto a solid substrate. In the gas phase, oxirane has been proposed to form by the dissociative recombination of C$_2$H$_5$O$^+$ with electrons,\cite{Dickens1997} where C$_2$H$_5$O$^+$ is formed by:

\begin{align}
    \ce{CH3+ + C2H5OH &-> C2H5O+ + CH4}\\
    \ce{H3O+ + C2H2 &-> C2H5O+ + h\nu}
\end{align}

The importance of this reaction has been difficult to quantify due to the lack of detections of ethanol, H$_3$O$^+$, C$_2$H$_2$, and CH$_3^+$ in cold cores, combined with uncertainties in the experimental branching ratios. Another plausible gas-phase source of oxirane at low temperature is the reaction between O atoms and the ethyl radical (C$_2$H$_5$), but the branching ratio compared to its more stable isomers, acetaldehyde and vinyl alcohol, is unknown. 

A reaction mechanism to form thiirane, S($^1$D) + C$_2$H$_4$, has been inferred by Balucani due to its similarity to N($^1$D) reactions.\cite{Balucani2009} The electrophilic S($^1$D) atom adds, without a barrier, to the double bond of ethylene,  forming the internally excited cyclic intermediate thiirane. Various ring-opening and isomerization pathways exist to convert thiirane into more stable products, and thus it is unclear if this reaction can produce thiirane under dense cloud conditions. Lastly, while this reaction has been found to be rapid down to 23 K, thiirane was not identified experimentally.\cite{Leonori2009}

Limited experimental and theoretical data exist for the formation of the 5-membered oxygen, nitrogen and sulfur heterocycles. Furan has been shown to form in the reaction between CH and acrolein at room temperature.\cite{Lockyear2013} The major products were found to be  1,3-butadienal (60 $\pm$ 12)\% and furan (17 $\pm$ 10)\%. Reactions of OH and NH with 1,3-butadiene have been proposed to form furan and pyrrole but more data is needed on the reaction rate constants and products at low temperature.\cite{Tuazon1999,McCarthy:2021aa}

The formation of the 6-membered nitrogen heterocycle, pyridine, in the interstellar medium has been an active area of research. Chemical models of Titan's atmosphere have proposed that pyridine may form by radical mediated reactions of hydrogen cyanide (HCN) with acetylene (C$_2$H$_2$);\cite{Ricca2001} however, this mechanism has not been experimentally verified and may not operate under dense cloud conditions. A similar mechanism involving sequential reactions of HCN with the acetylene radical cation has been shown experimentally to produce the pyrimidine ion.\cite{Hamid2014} In addition, recent ab initio molecular dynamics simulations and density functional theory computations have shown that pyridine ions can form during the ionization of van der Waals clusters of HCN and C$_2$H$_2$.\cite{Stein2020}

Formation of pyridine through ring expansion of pyrrole by methylidyne (CH) has been observed in the gas phase by Soorkia et al. (2010):\cite{Soorkia2010}

\begin{equation}
    \ce{CH + C4H5N -> c-C5H5N + H}.
\end{equation}

However, since pyrrole has likewise not been detected in the ISM, the formation of interstellar pyridine from pyrrole cannot be observationally constrained. Another neutral-neutral route to pyridine,  the reaction of CN with 1,3-butadiene, has been explored in a combined experimental and theoretical study.\citep{Morales:2011gl} The formation of the aliphatic product cyano-1,3-butadiene, however, was shown to be the major product.

\begin{align}
    \ce{CN + C4H6 &-> c-C5H5N + H}\\
    \ce{CN + C4H6 &-> CN-C4H5 + H}
\end{align}

The reaction of the cyano vinyl radical with vinyl cyanide is a promising route to pyridine at low temperature and has been verified experimentally, albeit at high temperatures in a pyrolytic reactor.\citep{Parker:2015iz}

\begin{equation}
    \ce{C2H3CN + C2H2CN -> C5H5N + CN}
\end{equation}

Supporting electronic structure calculations indicate that this reaction is exoergic and proceeds without an entrance barrier, therefore is likely to operate at low temperature. Future explorations of this reaction under low temperature thermal conditions are desirable, especially to determine the branching-ratio of pyridine and other potential products such as 1,4-dicyano-1,3-butadiene.

\citet{Fondren2007} studied the reaction of pyridine and pyrimidine with a series of ions in a selected ion flow tube. They inferred from these experiments that efficient ion-molecule radiative association reactions with HCN could form pyridine (and pyrimidine) from smaller ions:

\begin{align}
    \ce{C4H4+ + HCN &-> c-C5H5N+ + h\nu}\\
    \ce{C3H3N+ + HCN &-> c-C4H4N2+ + h\nu}
\end{align}

However, such radiative association reactions are challenging to measure in the laboratory and it remains unclear if the reactions above are efficient at low temperature. 
\citet{Balucani2019} investigated a pathway to pyridine via N($^2$D) reaction with C$_6$H$_6$, leading to a chain of unstable intermediate products that may decay to c-C$_5$H$_5$N. However, more recent computational work indicates that this reaction produces pyrrole, rather than pyridine, and that the major product channel is hydrogen cyanide plus a cyclopentadienyl radical ($\sim$90\%).\cite{Chin2021} 

\citet{Krasnopolsky2009} suggested another a neutral-neutral pathway to form pyridine; however, it is considered more likely that this reaction forms the aliphatic isomer C$_2$H$_5$C$_3$N:\cite{Loison2015}

\begin{equation}
    \ce{C3N + C2H6 -> c-C5H5N + H}
\end{equation}

N- and O-heterocyclic molecules, have been identified in the room-temperature residues resulting from the ultraviolet irradiation of the carbocyclic aromatic molecules benzene or naphthalene mixed in ices containing H$_2$O and NH$_3$.\cite{Materese2015} Nitrogen heterocycles identified include some of those searched for in our data: pyridine and quinoline. The oxygen-bearing heterocycles detected include the lactones pthalide (C$_8$H$_6$O$_2$), coumarin (C$_9$H$_6$O$_2$) and isocoumarin (C$_9$H$_6$O$_2$); however, it was noted that small O-heterocycles like furan could not be detected using their technique. 

\subsubsection{Formation of heterocyclic PAHs and nucleobases}

Like pure hydrocarbon PAHs, heterocyclic PAHs have been thought traditionally to form in high temperature environments through barriered reactions. One route suggested to operate in circumstellar envelopes is the reaction between meta-pyridyl radical with two acetylene molecules via a hydrogen abstraction/acetylene addition (HACA) type reaction mechanism.\cite{Parker2015} Following the formation of single-ring heterocyles, it has been suggested that formation of subsequent rings resulting in PANHs, PAOHs and PASHs should be efficient in the cold ISM. Pyridine has been suggested to be an important building block in the formation of more complex PANHs. Pyridine can be photolyzed to form pyridyl radicals (C$_5$H$_4$N),\cite{Lin2005} which have been shown to react without an entrance barrier with 1,3-butadiene to form the PANHs 1,4-dihydroquinoline and quinoline (as well as the iso- isomers), respectively.\cite{Parker2015,Parker2015b}

\citet{Ricca2001} proposed a route involving radical-mediated addition reactions of HCN and acetylene to form a pure aromatic ring. If hydrogen cyanide is abundant, the acetylene polymerization chemistry that leads to PAH formation in the circumstellar envelopes of AGB/post-AGB carbon stars can lead to the incorporation of nitrogen into the PAH backbone. They show that inclusion of a nitrogen atom into a PAH could promote the formation of additional hydrocarbon rings by lowering the ring closing barrier. Ultimately, these molecules could contribute to the formation of much larger PANHs that are believed to play an important role in interstellar chemistry.

The astrochemical link between simple, unsubstituted heterocycles and the more complex purine and pyrimidine nucleobases has not yet been made. \citet{Oba2019} have detected all three pyrimidine (cytosine, uracil and thymine) and three purine nucleobases (adenine, xanthine and hypoxanthine) in the organic residue of interstellar ices composed of  H$_2$O, CO, NH$_3$ and CH$_3$OH after exposure to UV photons. It is clear that heterocycles and even prebiotic nucleobases can form under conditions expected to be present in some regions of interstellar space, but the chemical mechanisms linking small molecules to the molecules of life remains unknown and will motivate future studies.

\subsubsection{Destruction of heterocycles and the effect on their detectability}

\citet{Peeters2005} compared the photolytic stabilities of benzene and pyridine (as well as pyrimidine and s-triazine) in matrix isolation experiments. Their results indicate N-heterocycles photolyse rapidly and their stability decreases with increasing number of N atoms in the ring. The authors calculated that these N-heterocycles in the gas phase would be destroyed in 10–100 years in the diffuse ISM, while in the Solar System 1 AU from the Sun they would not survive beyond several hours. They suggested that the only environment where small N-heterocycles could survive is in dense clouds that are shielded from intense UV-radiation. Pyridine and pyrimidine, but not the triply N-substituted ring s-triazine, could survive the average lifetime of a dense cloud. Photolytic stability data does not exist for the doubly N-substituted PANHs that we searched for here, so it is unclear whether molecules like the C$_8$H$_6$N$_2$ isomers could survive. Extrapolating from their data on pyrimidine, we expect that imidazole and pyridazine could likely survive destruction by UV photons in dense clouds, but more data is required to confirm this. It has not yet been reported how long these molecules and other heterocycles can survive in ices. In addition, the radiolytic stability of heterocycles to cosmic rays should be quantitatively investigated.  

Destruction of heterocycles through reactions are important to consider because they can potentially help explain their non-detection in TMC-1, but also, they can help determine molecules that may serve as detectable chemical proxies. Cyano-substituted aromatics have been shown to be good chemical proxies for small aromatics and PAHs due to the fact that CN radicals are likely to react without barriers with the double bonds of the aromatic ring.\cite{Lee:2019bc,Cooke:2020we} Because heterocycles typically lack the symmetry present in unsubstituted aromatics, the multiple CN-substitution sites decrease the chances of detecting a specific isomer. 

Recent theoretical work has suggested hydrogenated versions of heterocycles may serve as potential chemical proxies in dense clouds, due to the high H abundance, the ability of H to react via tunnelling and the efficient diffusion of H on interstellar dust grains.\cite{Miksch2021} The authors show that many heterocycles react slowly with H and suggest that hydrogenated derivatives of furan (2,3-dihydrofuran, and 2,5-dihydrofuran) and pyrrole (2,3-dihydropyrrole, 2,5-dihydropyrrole) are promising candidates for future interstellar searches.

\subsection{Suggested follow-up observations}

Our comparison of the upper limits derived from the single line and the MCMC analyses suggests that consideration of all lines within the spectral window of our broadband survey considerably constrains the upper limit determination. This suggests that the detectability of particular molecules depends not only on intrinsic molecular properties like the dipole moment (Figure \ref{fig:dipoles}), but also the spectral coverage and hence total integrated line intensity for a molecule. Broadband line surveys thus serve an even more essential role in the search for new species in the ISM. Observations at higher frequencies would be beneficial, as the brightest lines for most molecules lie outside of our GOTHAM spectral coverage. For example, the brightest lines are expected to fall between 100--200 GHz for the 3-membered rings, and $\sim$50--70 GHz for the 5- and 6-membered rings. 

In addition to broadband surveys as a tool to search for any heterocyclic species, targeted searches for chemically related molecules will serve an important role in constraining the chemistry of heterocycles in the ISM. As summarized in the preceding discussion, extensive experimental and theoretical studies have suggested plausible low-temperature heterocycle formation routes. While direct observation of the target heterocycles would provide the strongest support for proposed mechanisms, observations of the key reactive species en route to heterocycle formation can help to bolster or constrain these mechanisms. Some of these key molecules do not possess dipole moments (e.g. 1-3-butadiene), making their detection in TMC-1 challenging. Chemical proxies may be invoked, but only after careful validation through a combination of laboratory studies and kinetic models.

CN-substituted species have proven to be good choices as proxies, since the reaction between CN and unsaturated hydrocarbons is typically rapid,\cite{Sims1993} and CN imparts a strong dipole moment on the resulting reaction product(s). The use of benzonitrile as a proxy for benzene has been confirmed through laboratory measurements of CN + C$_6$H$_6$ down to temperatures close to those in TMC-1.\cite{Cooke:2020we} C$_2$H-substituted aromatic heterocycles may likewise serve as good chemical proxies. Ethynyl has been shown to react rapidly with unsaturated hydrocarbons,\cite{Chastaing:1998jp,Fortenberry_2021} including benzene,\cite{Goulay2006} and C$_2$H substituted cylic molecules, ethynyl cyclopentadiene and cyclopropenylidene, both of which have been recently detected toward TMC-1.\cite{Cernicharo2021,Cernicharo2021b}

Individual observations of key intermediate species in proposed chemical pathways will serve to motivate new experimental and theoretical studies and re-focus observational efforts. Direct observations in other sources of aromatic and/or heterocyclic solid-phase vibrational modes with the James Webb Space Telescope (JWST) would be particularly useful in constraining the gas-ice partitioning of these molecules.

\section{Conclusions}

In this study, we have applied a Bayesian inference analysis to the third data release of the GOTHAM collaboration survey of TMC-1 to constrain the upper limit column densities for a large selection of one- and two-ring heterocycles. For those few species previously searched for in TMC-1, our new analysis improves the upper limit constraint on these species. We compare our upper limits derived from the MCMC analysis with those derived from a single line analysis and find that, for all species, the consideration of all lines in the spectral window of the survey provides tighter constraints on the upper limit than consideration of only the single strongest line. In sum, our analysis confirms the peculiar depletion of heterocyclic species relative to the previously detected pure hydrocarbon cycles. Further improvements to our MCMC analysis, as well as complementary observations of chemically related species will help to further constrain the chemistry of heterocycles in the ISM.

\begin{acknowledgement}

This work was supported in part by funding from MIT, Union College and the University of British Columbia.  During portions of this work, funding for T.J.B. was provided through a Beckman Young Investigator Award made to B.A.M. by the Arnold and Mabel Beckman Foundation.
S.B.C. is supported by the NASA Astrobiology Institute through the Goddard Center for Astrobiology. The National Radio Astronomy Observatory is a facility of the National Science Foundation operated under cooperative agreement by Associated Universities, Inc. The Green Bank Observatory is a facility of the National Science Foundation operated under cooperative agreement by Associated Universities, Inc
\end{acknowledgement}

\begin{suppinfo}

Corner plots of column densities for all heterocycles investigated in this work.

\end{suppinfo}


\providecommand{\latin}[1]{#1}
\makeatletter
\providecommand{\doi}
  {\begingroup\let\do\@makeother\dospecials
  \catcode`\{=1 \catcode`\}=2 \doi@aux}
\providecommand{\doi@aux}[1]{\endgroup\texttt{#1}}
\makeatother
\providecommand*\mcitethebibliography{\thebibliography}
\csname @ifundefined\endcsname{endmcitethebibliography}
  {\let\endmcitethebibliography\endthebibliography}{}

\end{document}